\documentclass[aps,prb,twocolumn,showpacs,floatfix,superscriptaddress,footinbib]{revtex4}
\usepackage{amsmath,amssymb,natbib,bm,graphicx,url,epsfig,color,verbatim,psfrag,epstopdf}
\usepackage[ansinew]{inputenc}

\begin{document}

\title{Transport through a quantum dot with two parallel Luttinger liquid leads}

\author{Florian Elste}
\affiliation{Department of Physics, Columbia University, 538 West 120th Street, New York, NY 10027, USA}

\author{David R. Reichman}
\affiliation{Department of Chemistry, Columbia University, 3000 Broadway, New York, NY 10027, USA}

\author{Andrew J. Millis}
\affiliation{Department of Physics, Columbia University, 538 West 120th Street, New York, NY 10027, USA}

\date{\today}

\begin{abstract}

We study transport through a quantum dot side-coupled to two parallel Luttinger liquid leads in the presence of a Coulombic dot-lead interaction. This geometry enables an exact treatment of the inter-lead Coulomb interactions. We find that for dots symmetrically disposed between the two leads the correlation of charge fluctuations between the two leads can lead to  an enhancement of the current at the Coulomb-blockade edge and even to a negative differential conductance. Moving the dot off center or separating the wires further converts the enhancement to a suppression.

\end{abstract}


\maketitle

\section{Introduction} 

Quantum dots and single-molecule devices reveal many  quantum phenomena  of fundamental and perhaps practical interest.\cite{Park,Jo,Galperin} However, the role of a Coulombic interaction between the charges on the dot and on the leads 
has been little studied, even though this Coulomb interaction is clearly relevant. Dot-lead interactions were studied by Boulat and Saleur\cite{Boulat08a}, and Boulat, Saleur and Schmitteckert\cite{Boulat08b} in the context of an interacting resonant-level model. Goldstein and collaborators presented general duality-based results that they noted were applicable also to models with dot-lead interactions.\cite{Goldstein09} In previous papers we considered different aspects of quantum dots Coulomb-coupled to leads, including the relaxational dynamics of a quantum dot side-coupled to a single Luttinger liquid lead in equilibrium  (in Ref.~\onlinecite{Elste1}) and the case of a quantum dot coupled to two one-dimensional leads subject to a finite bias voltage (in Ref.~\onlinecite{Elste2}). However, in the theoretical work to date \textit{inter}-lead couplings have not been considered. While inter-lead interactions may be negligible if the screening of the Coulomb potential is strong, in general one expects molecular junctions to involve quantum dots which are very close to the leads, so that the inter-lead interaction is not negligible.  In this paper we show that this interaction can be of crucial importance, because it suppresses fluctuations which act to decohere the different states.  Thus, electron-electron interactions can enhance transport, leading (in appropriate circumstances which we define below) to conductances which peak at the Coulomb-blockade edge and thus exhibit a negative differential conductance. 

\begin{figure}[!t]
\begin{center}
\includegraphics[width=7.5cm,angle=0]{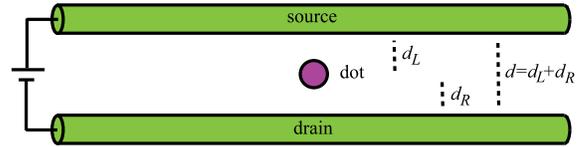}
\caption{Schematic of a one-dimensional lead-dot-lead system showing two parallel leads.}\label{model_fig} 
\end{center}
\end{figure}

A lead may be modelled as a quantum wire described by Luttinger liquid physics.\cite{Fabrizio94,Maurey,Lerner} However, for general geometries, inter-lead interactions do not fit easily into the Luttinger liquid formalism, because the boundary breaks translational invariance and the interactions give rise to a complicated boundary condition on the charge modes. We are not   aware of a convincing theoretical treatment of this boundary condition. In this paper we therefore specialize to the geometry shown in Fig.~\ref{model_fig}, for which the inter-lead interaction may be treated by standard Luttinger liquid methods. This geometry reveals the essential physics of the inter-lead coupling. 

The rest of this paper is organized as follows. Section \ref{Model} defines the model we study and the methods we use. Section \ref{electronic_tunneling} gives the basic theoretical results, Sec.~\ref{current_voltage_characteristics} presents the experimental consequences, and Sec.~\ref{conclusions} is a conclusion.

\section{Model and Methods\label{Model}}

\subsection{Model}

The quantum dot problem sketched in Fig.~\ref{model_fig} is described by a Hamiltonian of the general form 
\begin{equation}
H = H_\text{lead} + H_\text{dot} +  H_\text{Coul}+H_\text{mix}.
\label{H}
\end{equation}
We label the two leads by $\alpha=L,R$, the lead orbitals by $a$ and write the lead Hamiltonian as
\begin{align}
H_\text{lead} & ~=~ \sum_{\alpha=L,R} \sum_{a k \sigma} \epsilon^\alpha_{k} c^{\dagger}_{\alpha a k \sigma} c_{\alpha a k \sigma} \nonumber \\
~+ \frac{1}{2} & \int dx \, dx' 
\sum_{\alpha\alpha'=L,R} \sum_{aa'} V_{\alpha\alpha'}(x-x') {\rho}_{\alpha a}(x){\rho}_{\alpha' a'}(x').
\label{Hlead}
\end{align}
The operator $c^{\dagger}_{\alpha a k \sigma}$ creates an electron with momentum $k$, energy $\epsilon^\alpha_k$, and spin $\sigma$ in state $a$ of lead $\alpha$. The interaction $V$  depends on the difference in position $x-x{'}$ and is the sum of two terms, an intra-lead interaction
\begin{equation}
V_{\alpha\alpha}(x-x') = \frac{e^2}{\epsilon|x-x'|}
\end{equation}
(already considered in Ref.~\onlinecite{Elste2}), and an inter-lead interaction denoted by
\begin{equation}
\quad V_{\alpha \bar{\alpha}}(x-x') = \frac{e^2}{\epsilon\sqrt{(x-x')^2+d^2}}
\end{equation}
with $\bar{\alpha}=L$ if $\alpha =R$ and vice versa. $\rho_{\alpha a}(x)$ is the corresponding operator giving the charge density at position $x$, $\epsilon$ is a background dielectric constant, and $d$ denotes the spacing between the leads.  The convenient feature of the geometry we consider is that both intra- and inter-lead interactions are functions only of the relative distance along the lead, permitting use of usual bosonisation techniques. 

The quantum dot Hamiltonian $H_\text{dot}$ may be written as 
\begin{equation}
H_\text{dot}= \varepsilon_d \, n_d +\frac{U}{2} n_d(n_d-1),
\label{Hdot}
\end{equation}
where $U$ is the dot charging energy,
$n_d=\sum_{\sigma} d_{\sigma}^\dagger d_{\sigma}$ is the total dot density,
and $d_{\sigma}^\dagger$ creates an electron with energy $\varepsilon_d$ and spin $\sigma$ on the dot.

The Coulomb interaction between electrons on the quantum dot and electrons in the leads 
may be written as
\begin{equation}
H_\text{Coul} = n_d \sum_{\alpha=L,R} \sum_{a}\int d x \, W_\alpha(x) {\rho}_{\alpha a}(x)
\label{Hcoul}
\end{equation}
with 
\begin{equation}
W_\alpha(x) = \frac{e^2}{\epsilon \sqrt{x^2+d_\alpha^2}}
\end{equation}
and $d_\alpha$ the distance between the quantum dot and lead $\alpha$.
The dot-lead hybridization is given by 
\begin{equation}
H_\text{mix} = \sum_{\alpha=L,R} \sum_{a k \sigma} \left[ \mathcal{T}_\alpha \, d^{\dagger}_{\sigma} c_{\alpha a k \sigma} + \mathcal{T}^*_\alpha \, c^{\dagger}_{\alpha a k \sigma} d_{\sigma} \right].
\end{equation}

\subsection{Bosonisation}
 
We bosonise the system as in our previous papers;\cite{Elste1,Elste2} the new feature is the inter-lead coupling, which mixes the boson modes of the two leads.
We consider a system with linear dimension $L$ and periodic boundary conditions, and
combine spin and orbital quantum number into a superindex $\beta=1,\dots,M$. 
The physics is represented in terms of right ($\lambda=+$) and left ($\lambda=-$) moving particle-hole pairs,\cite{Haldane81}
which can be recombined into boson operators 
${\phi}_{\alpha\beta}(q)$, $\Pi_{\alpha\beta}(q)$ 
that obey the volume commutation relation
$[\phi_{\alpha\beta}(q),\Pi_{\alpha'\beta'}(-q')]=i(L/\pi) \delta_{\alpha\alpha'} \delta_{\beta\beta'} \delta_{qq'}$.
The total particle density in lead $\alpha$ and state $\beta$ is given by
$\rho_{\alpha\beta}(q)=iq\phi_{\alpha\beta}(q)$.

The lead electron creation operator $\psi_{\alpha\lambda\beta}$ may be rewritten in terms of the bosons as\cite{Luther,Giamarchi}
\begin{equation}
\psi_{\alpha \lambda \beta}(x)=\frac{U_{\alpha\lambda}}{\sqrt{2\pi\eta}}e^{i\lambda k_F x}e^{ i \frac{\pi}{L}\sum_q e^{iqx} \left[ \lambda{\phi}_{\alpha\beta}(q)-\frac{1}{iq} \Pi_{\alpha\beta}(q)
\right]}.
\label{psidef}
\end{equation}
Here the small positive infinitesimal factor $\eta$ arises from the correct normal ordering of the operators.
The operator $U_{\alpha\lambda}$ denotes the Klein factor which carries the Fermi statistics.
Left and right lead electrons are assumed to be at different
chemical potentials $\mu_\alpha$, which gives rise to a different time evolution
of the Klein factors, $U_{\alpha\lambda}(t) = e^{\mu_\alpha t} U_{\alpha\lambda}(0)$.

The symmetry between the two leads means that their  low-energy physics  can be described in terms of a set of new boson operators $\phi_{\alpha b},\Pi_{\alpha b}$ with a parity quantum number $p=\text{e},\text{o}$ as well as a mode label $b$. The new operators are related   by a linear transformation to the $\phi_{\alpha\beta},\Pi_{\alpha\beta}$ defined in terms of the fermions. In  terms of the new operators the lead Hamiltonian becomes 
\begin{align}\label{Hamiltonian3}
H_\text{lead} & ~=~ \sum_{p= \text{e},\text{o}} \sum_{b=1,\dots,M} \sum_q \frac{\pi}{2 L} v_{p b}(q) \nonumber \\
\times \bigg[ & K_{p b}(q) \, \Pi_{p  b}(-q) \Pi_{p  b}(q) + \frac{q^2}{K_{p b}(q)}\phi_{p  b}(-q) \phi_{p  b}(q) \bigg]
\end{align}
with even ($\text{e}$) and odd ($\text{o}$) boson modes defined by 
\begin{equation}
\Pi_{\text{e} b}(q),\Pi_{\text{o} b}(q) = \frac{\Pi_{L b}(q) \pm \Pi_{R b}(q)}{\sqrt{2}},
\end{equation}
and
\begin{equation}
\phi_{\text{e} b}(q),\phi_{\text{o} b}(q) = \frac{\phi_{L b}(q) \pm \phi_{R b}(q)}{\sqrt{2}},
\end{equation}
and Luttinger parameters $v_{\text{e} b}$, $v_{\text{o} b}$ and $K_{\text{e} b}$, $K_{\text{o} b}$ determined by the bare velocities and interactions of the lead eigenstates.
In terms of these boson modes, the electron operator becomes

\begin{widetext}
\begin{align}
\psi_{L \lambda \beta}(x) & ~=~ U_{L\lambda} e^{i\frac{\pi}{\sqrt{M}L} \sum_q e^{iqx} \frac{1}{\sqrt{2}} \left[ \lambda ({\phi}_{\text{e},b=1}(q) + {\phi}_{\text{o},b=1}(q))-\frac{1}{iq} (\Pi_{\text{e},b=1}(q)+\Pi_{\text{o},b=1}(q))\right]} \, \psi_{L \lambda \beta}^\text{rest}(x), \\
\psi_{R \lambda  \beta}(x) & ~=~ U_{R\lambda} e^{i\frac{\pi}{\sqrt{M}L} \sum_q e^{iqx} \frac{1}{\sqrt{2}} \left[ \lambda ({\phi}_{\text{e},b=1}(q)-{\phi}_{\text{o},b=1}(q))-\frac{1}{iq} (\Pi_{\text{e},b=1}(q)-\Pi_{\text{o},b=1}(q))\right]} \, \psi_{R \lambda \beta}^\text{rest}(x).
\end{align}
\end{widetext}
with
$\psi_{\alpha\lambda \beta}^\text{rest}$ an exponential of a combination of the
${\phi}_{\alpha,b}$, $\Pi_{\alpha,b}$ with $b\geq 2$

Rewriting the Coulomb potential as 
\begin{equation}
\frac{e^2}{|x|}=\frac{e^2}{L}\sum_qe^{iqx} V_{\alpha\alpha}(q), \quad V_{\alpha\alpha}(q) = \log \left( 1+ \frac{1}{\Lambda^2 q^2}\right)
\end{equation}
with $\Lambda$ a short-distance cutoff and 
\begin{equation}\label{define_log}
\frac{1}{\sqrt{x^2+d^2}}=\frac{1}{L}\sum_qe^{iqx} V_{\alpha\bar{\alpha}}(q), \quad V_{\alpha\bar{\alpha}}(q) = \log \left( 1+ \frac{1}{d^2 q^2}\right)
\end{equation}
we find for the Luttinger parameters in Eq.~(\ref{Hamiltonian3}) that
\begin{eqnarray} 
K_{\text{e} 1}(q),K_{\text{o} 1}(q)&=&\frac{1}{\sqrt{1+MV_c[V_{\alpha\alpha}(q)\pm V_{\alpha\bar{\alpha}}(q)]}},
\label{K1def}\\
v_{\text{e} 1}(q),v_{\text{o} 1}(q)&=&v_F\sqrt{1+MV_c[V_{\alpha\alpha}(q)\pm V_{\alpha\bar{\alpha}}(q)]}\label{v1def}
\end{eqnarray}
and $K_{\text{e} b}(q)=K_{\text{o} b}(q)=1$, $v_{\text{e} b}(q)=v_{\text{o} b}(q)=v_F$ for $b>1$.
Here we introduced a measure of the Coulomb interaction strength 
and defined the dimensionless parameter $V_c=e^2/\pi v_F\epsilon$.

\subsection{Canonical transformation}

To treat the dot-lead interaction we use the canonical transformation methods of Refs.~\onlinecite{Elste1,Elste2}. Following the approach of these papers but  expressing the dot-lead interaction in terms of even and odd boson fields yields
\begin{align}
H_\text{Coul} &= \sqrt{\frac{M}{2}} \frac{\pi v_F V_c}{L} \sum_q iq \bigg( \phi_{\text{e} 1}(q) \left[ W_{L}(q) +W_{R}(q) \right] \nonumber \\
&~+ \phi_{\text{o} 1}(q) \left[  W_{L}(q) - W_{R}(q) \right] \bigg) \, n_d 
\label{Hcoul1d}
\end{align}
with
$W_{\alpha}(q) = \log ( 1+ 1/d_\alpha^2 q^2)$.
We observe that for a  dot placed symmetrically between the two leads, the coupling to the odd mode vanishes exactly.

The linear coupling between dot occupancy $n_d$ and lead density may now be removed by a canonical transformation \cite{Schotte} which shifts 
\begin{equation}
\phi_{\alpha 1}(q) \rightarrow \phi_{\alpha 1}(q)-Z_{\alpha}(q)n_d/iq\sqrt{2M}
\end{equation}
with 
\begin{align}
Z_{\text{e}}(q) & = \frac{v_FK_{\text{e}1}(q)}{v_{\text{e}1}(q)}MV_c \left[ W_{L}(q) + W_{R}(q) \right] \\
Z_{\text{o}}(q) & = \frac{v_FK_{\text{o}1}(q)}{v_{\text{o}1}(q)}MV_c \left[ W_{L}(q) - W_{R}(q) \right].\label{Zdef}
\end{align}
The canonical transformation acts on an operator ${\cal O}$ by ${\cal O}\rightarrow e^{iS}{\cal O}e^{-iS}$ with
\begin{equation}
S = - n_d \frac{\pi}{\sqrt{2M}L}\sum_{p=\text{e},\text{o}} \sum_q Z_{\alpha}(q) \frac{\Pi_{\alpha 1}(-q)}{iq}.
\label{Sdef}
\end{equation}
Under the canonical transformation the dot operator becomes
\begin{equation}
d^\dagger_{\sigma}\rightarrow d^\dagger_{\sigma} e^{\frac{i \pi}{\sqrt{2M}L} \sum_q \frac{1}{iq} \left[ \Pi_{\text{e} 1}(q) Z_{\text{e}}(q) + \Pi_{\text{o} 1}(q) Z_{\text{o}}(q)\right]}.
\end{equation}

\section{Electronic tunneling}\label{electronic_tunneling}

The crucial quantities in our considerations are expectation values of the form
\begin{equation}\label{definecorrelator}
F_{\alpha}(t)=\left\langle 
\xi^\dagger_{\alpha \lambda \beta}(t) \xi_{\alpha \lambda \beta}(0)
\right\rangle
\end{equation}
with $\sum_{\alpha \lambda \beta} \mathcal{T}_\alpha \xi_{\alpha \lambda \beta}+\text{h.c.}$ the renormalized hybridization Hamiltonian $e^{iS}H_\text{mix}e^{iS}$ without the $d_\sigma$ operator.
We obtain, say, for $\alpha=L$
\begin{equation}
F_L(t)=F_0(t)e^{\Phi_L(t)}e^{-i\mu_L t}
\end{equation}
with $F_0(t)$ the free-fermion correlation and
\begin{align}
\Phi_L & (t) ~=~ \frac{2\pi}{ML}\sum_{q\neq 0}\frac{1}{2|q|} \bigg[
1-e^{-i v_F|q| t} \nonumber \\
- & B_{\text{e}}(q)\left(1-e^{-i v_{\text{e}1}(q)|q| t}\right) 
-B_{\text{o}}(q)\left(1-e^{-i v_{\text{o}1}(q)|q| t}\right) \bigg],
\end{align}
where we have defined
\begin{align}
B_{\text{e}}(q),B_{\text{o}}(q) & ~=~ \frac{1}{4} \bigg( K_{\text{e}1}(q)+ \frac{\left[1-Z_{\text{e}}(q)\right]^2}{K_{\text{e}1}(q)} \nonumber \\
& ~~~~~~~~+ K_{\text{o}1}(q)+ \frac{\left[1 \mp Z_{\text{o}}(q)\right]^2}{K_{\text{o}1}(q)} \bigg).
\end{align}
To obtain an idea of the effects of screening, we approximate the 
logarithmic functions $K_{\text{e}1}(q)$, $K_{\text{o}1}(q)$ and $v_{\text{e}1}(q)$, $v_{\text{o}1}(q)$ by constants 
$K_{\text{e}1}$, $K_{\text{o}1}$ and $v_{\text{e}1}$, $v_{\text{o}1}$. In this case the decay of electronic correlations
is described by 
\begin{equation}
F_\alpha(t) \propto \frac{e^{-i\mu_\alpha t}}{\left[ \frac{v_F}{\pi T \Lambda}\sinh (\pi T t) \right]^{Y_\alpha}}
\end{equation}
with the Luttinger exponents
\begin{equation}\label{define_YL}
Y_L, Y_R = \frac{1}{M} \bigg ( M-1+ \frac{K_{\text{e} 1}+\frac{\left[1-Z_{\text{e}}\right]^2}{K_{\text{e}1}}+K_{\text{o}1}+\frac{\left[1\mp Z_{\text{o}}\right]^2}{K_{\text{o}1}}}{4} \bigg).
\end{equation}
This formula plays a  crucial role in our subsequent analysis.  We see that the odd-parity channel shift $Z_\text{o}$ (whose sign depends on which lead is nearer to the dot) enters with opposite sign in the two exponents. 

\begin{figure}[!t]
\begin{center}
$\begin{array}{c}
\textbf{(a)} \\ 
\includegraphics[width=7.8cm,angle=0]{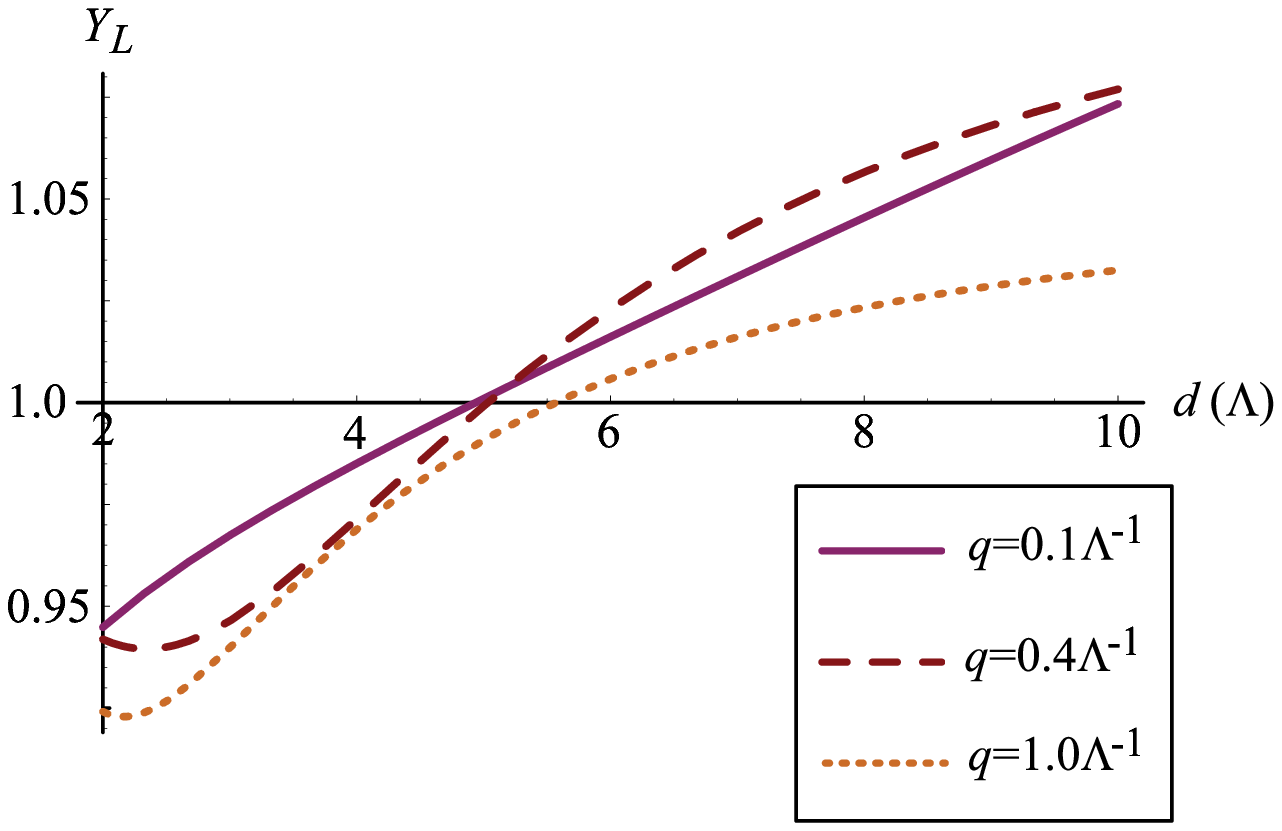} \\
\textbf{(b)} \\ 
\includegraphics[width=7.8cm,angle=0]{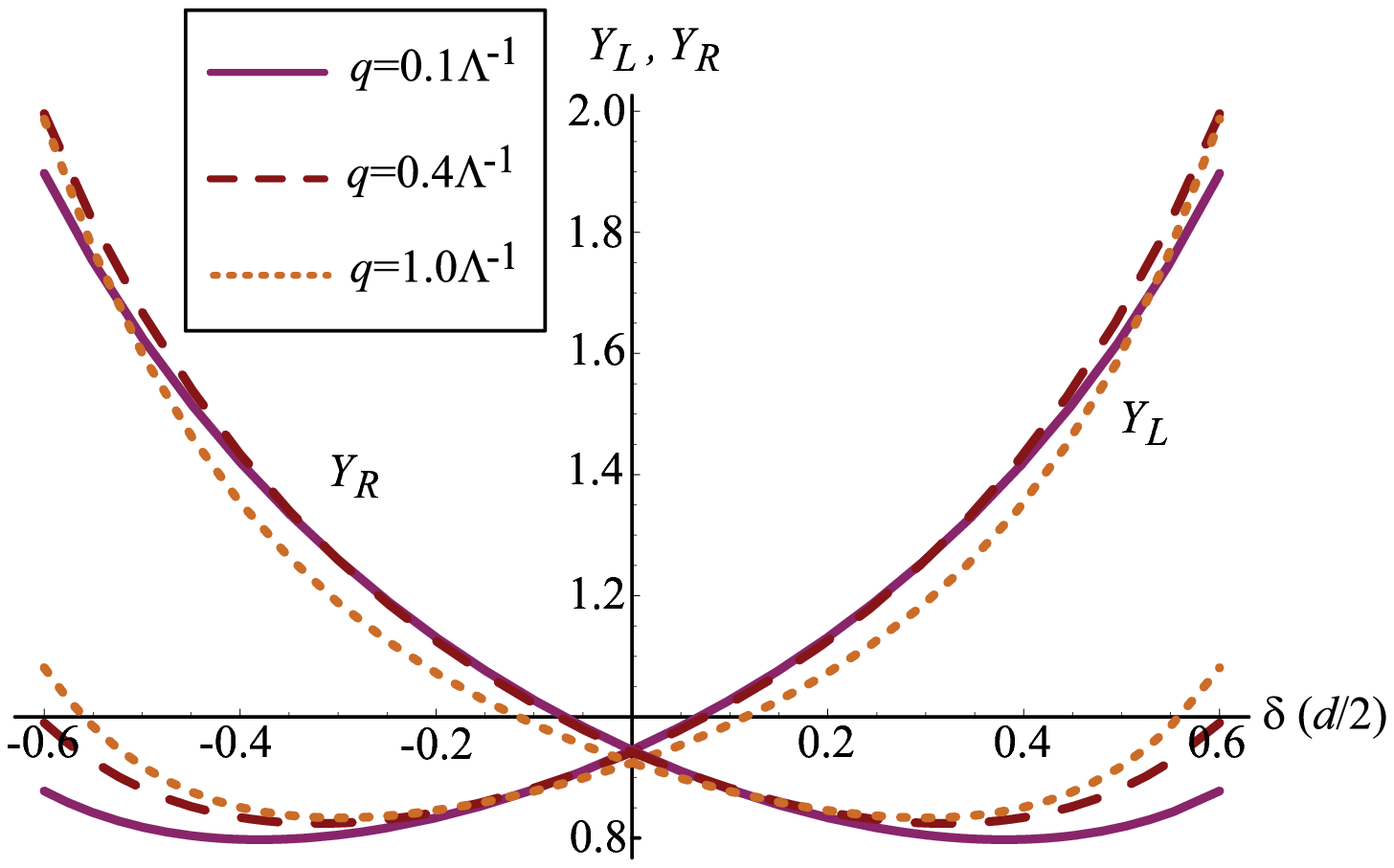} 
\end{array}$
\caption{(a) Luttinger exponent $Y_L$ of a symmetric junction ($Y_R=Y_L$) with nanotube leads as a function of the inter-lead distance $d$ for different $q$ values.
(b) Luttinger exponents $Y_L,Y_R$ as a function of the displacement $\delta\equiv d_L-d/2$ of the quantum dot relative to the leads. [Here $\delta=0$ represents a symmetric junction.}\label{exponents} 
\end{center}
\end{figure}

Equation~(\ref{define_YL}) implies a strong effect of the geometry on the electronic tunneling. In the symmetric case with equal dot-lead distances $d_L=d_R=d/2$,  we see $Z_{\text{o}}=0$, because the odd boson mode does not contribute to the dot-lead coupling in that case. As the dot is moved off center, one of the exponents increases and the other decreases; we shall see that this has important consequences for the tunneling. The dependence of the exponents on inter-lead distance and on the relative position of the lead is shown in Fig.~\ref{exponents}. 

It is interesting to compare Eq.~(\ref{define_YL}) to the expressions obtained for the models studied in our previous work.\cite{Elste1,Elste2}
Reference~\onlinecite{Elste1} considered a dot coupled to a single Luttinger liquid. In this case we have only one channel, and we found
\begin{equation}
Y = \frac{1}{M} \left( M-1+ \frac{K+\frac{\left(1-Z\right)^2}{K}}{2} \right),
\end{equation}
which can be obtained from our Eq.~(\ref{define_YL}) by  dropping, say, all of the right channel interactions and couplings, which in practice means replacing $Z_{\text{e}}$ and $Z_{\text{o}}$ by $Z$, and $K_{\text{e}1}$ and $K_{\text{o}1}$ by $K$. The physics is therefore relatively similar between the two cases, except that because the odd-channel interactions are weaker than the even channel ones, the exponent can be a little closer to the non-interacting value.

However, if the quantum dot is coupled to two leads but we ignore the  inter-lead electron-electron interactions, we may replace $Z_{\text{e}}$ by $Z_L+Z_R$, $Z_{\text{o}}$ by 
$Z_L-Z_R$ and $K_{\text{e}1}$, $K_{\text{o}1}$ by $K$ 
to obtain
\begin{equation}
Y_L = \frac{1}{M} \left( M-1+ \frac{K+\frac{\left(1-Z_L\right)^2}{K}+\frac{Z_R^2}{K}}{2} \right),
\end{equation}
which is consistent with our result for the Luttinger exponent obtained in Ref.~\onlinecite{Elste2}, except that in this paper boundary rather than bulk exponents appeared. The crucial differences are the factor of $2$ in the denominator and the factor of $Z_R^2/K$ which represented the 'orthogonality' effect caused on one lead by adding a charge from the other. This orthogonality effect acts to suppress electron transport. By correlating the fluctuations in the two leads, the inter-lead Coulomb interaction acts to suppress this orthogonality effect; we shall see that the result is an enhanced tunneling. 

These arguments are supported by Fig.~\ref{exponents}. The upper panel shows that the exponent increases with increasing inter-lead distance $d$, reflecting the weakening of the inter-lead interaction, the resulting enhancement of relative fluctuations, and thus an increasing orthogonality effect. The lower panel shows that there is a relative dot-lead distance at which the exponent for tunneling from a given lead is minimized, reflecting the interplay between the increase in $Z_\text{o}$ and the decrease in $Z_\text{e}$ due to the weakening of one of the couplings. For a very asymmetric situation we revert to the single-lead case studied previously.

\section{Current-voltage characteristics}\label{current_voltage_characteristics}

The assumption of weak hybridization allows us to 
investigate the nonequilibrium dynamics of the system
using a master-equation approach.
The quantum dot with a spin-degenerate level is described by the diagonal density matrix, 
which contains the occupation probabilities $\mathcal{P}_0$, $\mathcal{P}_1$, and $\mathcal{P}_2$ of the empty state $|0\rangle$, the singly-charged state $|1 \rangle $, and the doubly-charged state $|2 \rangle $. 
Expanding the von Neumann equations to lowest order in the hybridization 
and making use of the Markov approximation yields a set of master equations 
as discussed in Refs.~\onlinecite{Elste1,Elste2}
with the tunneling rates related to the correlation function in Eq.~(\ref{definecorrelator}).
For example, the rate to move an electron from lead $\alpha$ to the empty
dot is given by 
$\mathcal{R}^{\alpha}_{0 \rightarrow 1} = 2 \, \text{Re} \int_0^\infty d\tau F_\alpha(\tau) \, e^{-i\tilde{\varepsilon}_d\tau}$.
Evaluating the expression we find that
\begin{equation}
\mathcal{R}^{\alpha}_{0 \rightarrow 1} \propto
\frac{1}{\tau_\alpha} \, \left(\frac{|\tilde{\varepsilon}_d-\mu_\alpha|}{v_c\Lambda}\right)^{Y_\alpha-1} \, \theta\left( \mu_\alpha-\tilde{\varepsilon}_d \right)
\label{T0rate}
\end{equation}
in the zero-temperature limit.
Here $1/\tau_\alpha \propto 2\pi |\mathcal{T}_\alpha|^2 / (v_{\text{e}}/\Lambda)$ is the bare tunneling rate, and the Luttinger parameters have been approximated by constants.

\begin{figure}[!t]
\begin{center}
$\begin{array}{c}
\textbf{(a)} \\
\includegraphics[width=7.8cm,angle=0]{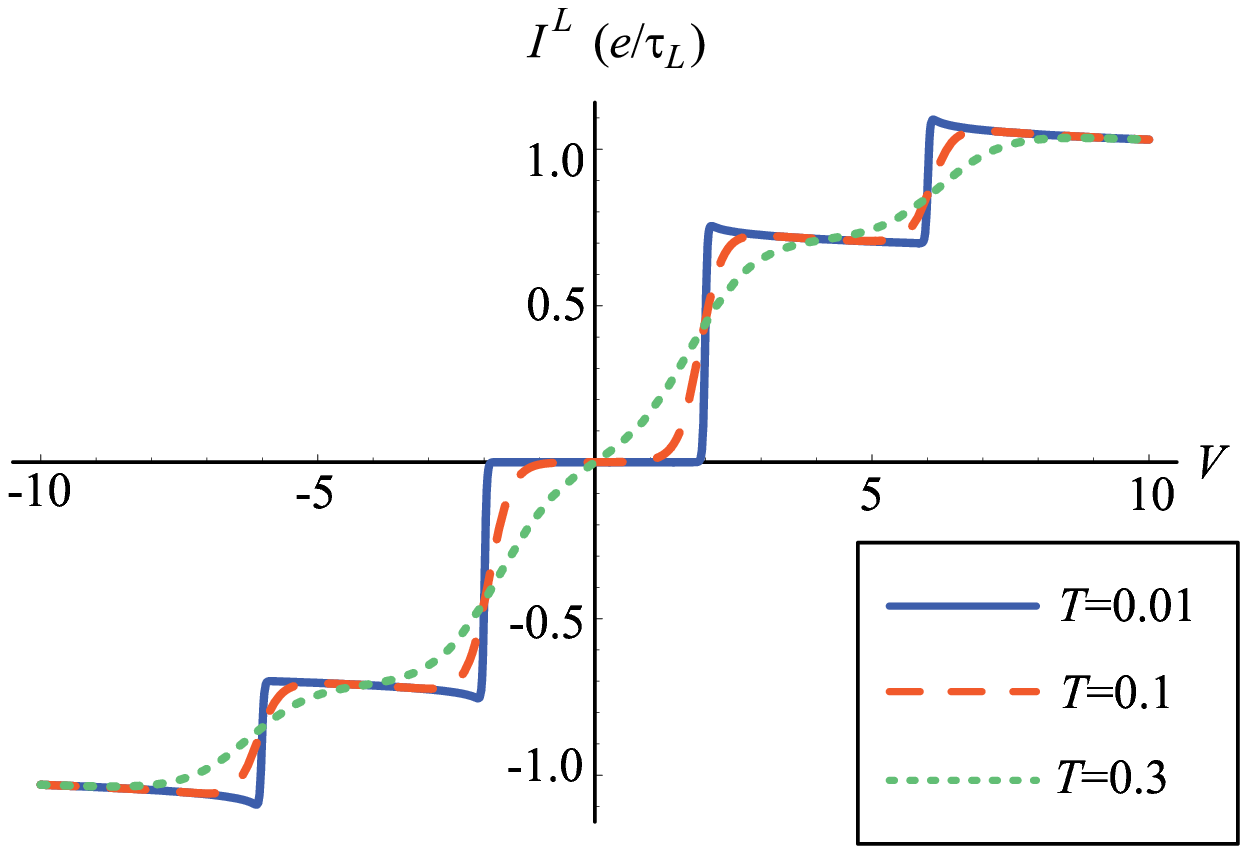} \\
\textbf{(b)} \\
\includegraphics[width=7.8cm,angle=0]{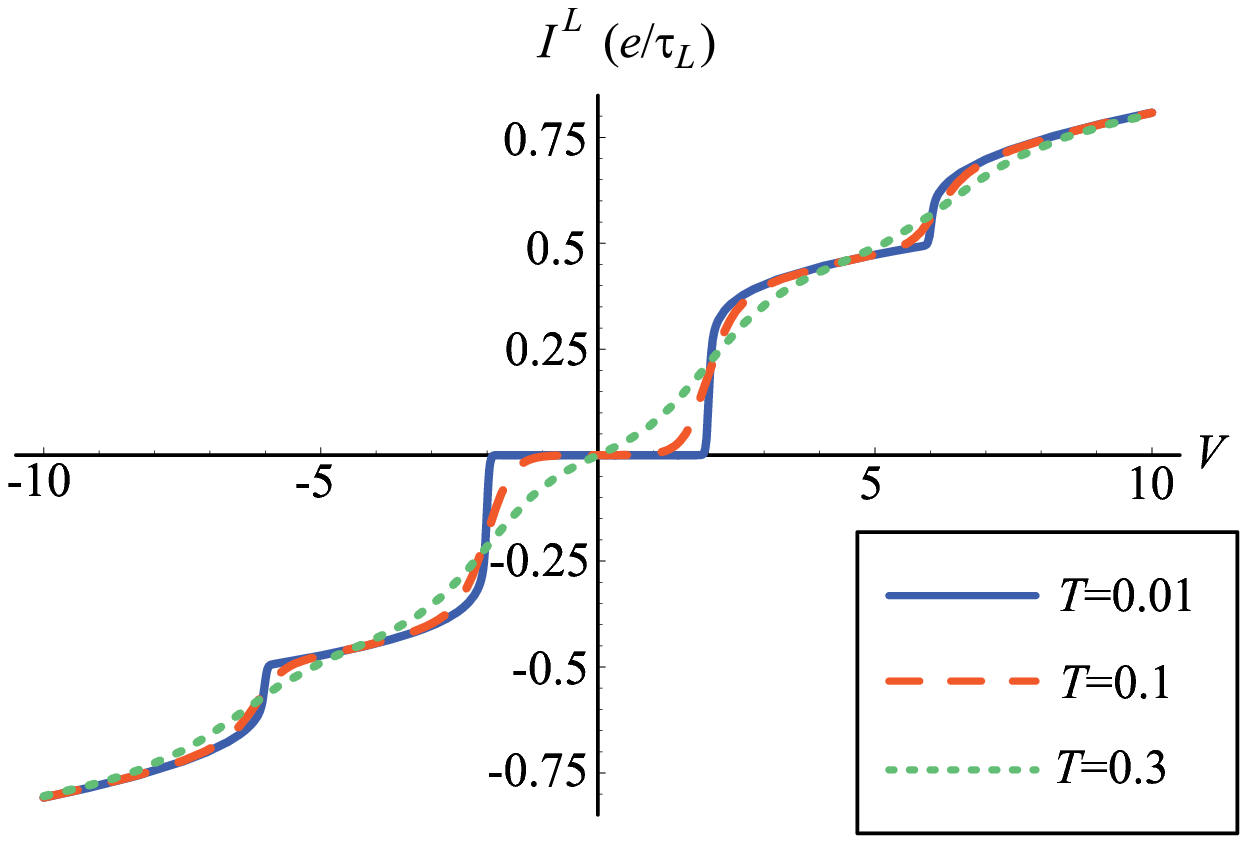}
\end{array}$
\caption{Comparison of current-voltage characteristics of a 
quantum dot symmetrically side-coupled to two parallel Luttinger liquid leads (a) 
for a model including inter-lead Coulomb interactions and (b) for the same model but
with inter-lead Coulomb interactions set to zero.
Voltages and thermal energies are given in units of the corresponding onsite energy. 
For the results shown in (a), the interaction parameters are obtained from Eq.~(\ref{define_YL})
using Eqs.~(\ref{K1def})---(\ref{v1def}) and (\ref{Zdef}) with $V_c\simeq 0.9$, $M=4$ (channels), $d_\alpha=d/2=\Lambda$,
and with the logarithms evaluated at $q=0.1\, \Lambda^{-1}$ (these parameters were shown in Ref.~[\onlinecite{Elste2}] to be appropriate for a carbon nanotube with $\Lambda$ of the order of the tube diameter).
The results shown in (b) are obtained for the same parameters except  $V_{\alpha\bar{\alpha}}$ [cf.~Eq.~(\ref{define_log})] set to zero.
The local Coulomb interaction is assumed to be twice as large as the onsite energy in both panels.}\label{IVcurves} 
\end{center}
\end{figure}

Solving the set of master equations in the steady state ($\dot{\mathcal{P}}_i=0$)
gives an expression for the steady-state current in terms of the transition rates. In the limit of very large $U$, the rates
for the excitation of two electrons on the dot vanish and the current is described 
by the simple expression
\begin{equation}\label{analytic_ex}
\langle I_{\alpha} \rangle =
2e\, \frac{\mathcal{R}^{\alpha}_{0 \rightarrow 1} \mathcal{R}^{\bar{\alpha}}_{1 \rightarrow 0} - \mathcal{R}^{\alpha}_{1 \rightarrow 0} \mathcal{R}^{\bar{\alpha}}_{0 \rightarrow 1}}{2 \mathcal{R}_{0 \rightarrow 1} + \mathcal{R}_{1 \rightarrow 0}} \nonumber \end{equation}
with $\mathcal{R}_{n \rightarrow m}\equiv\sum_\alpha \mathcal{R}^\alpha_{n \rightarrow m}$.
Our numerical results are obtained with the full expression, which is 
 presented in Ref.~\onlinecite{Elste2}. Also in our numerical results we used the parameters identified in Ref.~\onlinecite{Elste2} as corresponding to a nanotube with the Coulomb interaction cut off at a momentum scale of one tenth of the inverse tube diameter. 

\begin{figure}[!t]
\begin{center}
$\begin{array}{c}
\textbf{(a)} \\ 
\includegraphics[width=7.8cm,angle=0]{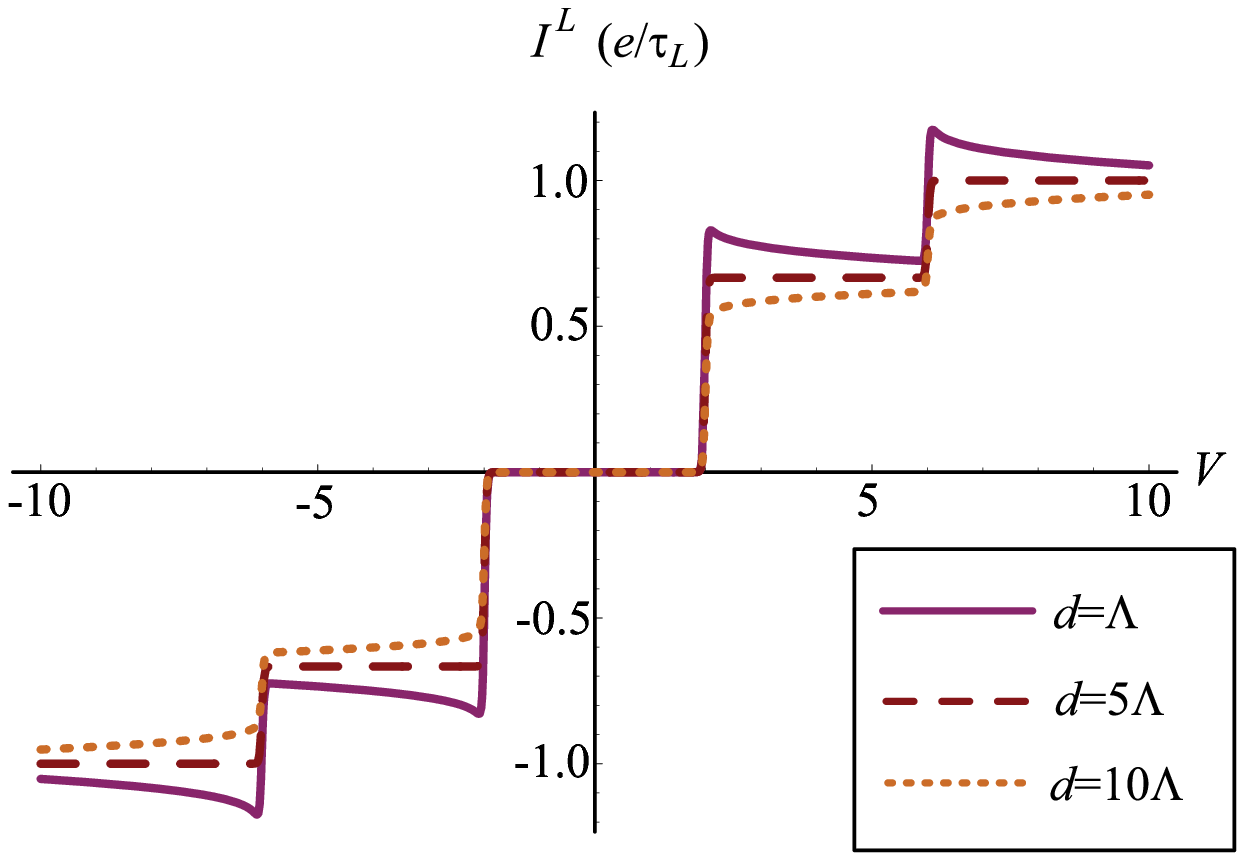} \\
\textbf{(b)} \\ 
\includegraphics[width=7.8cm,angle=0]{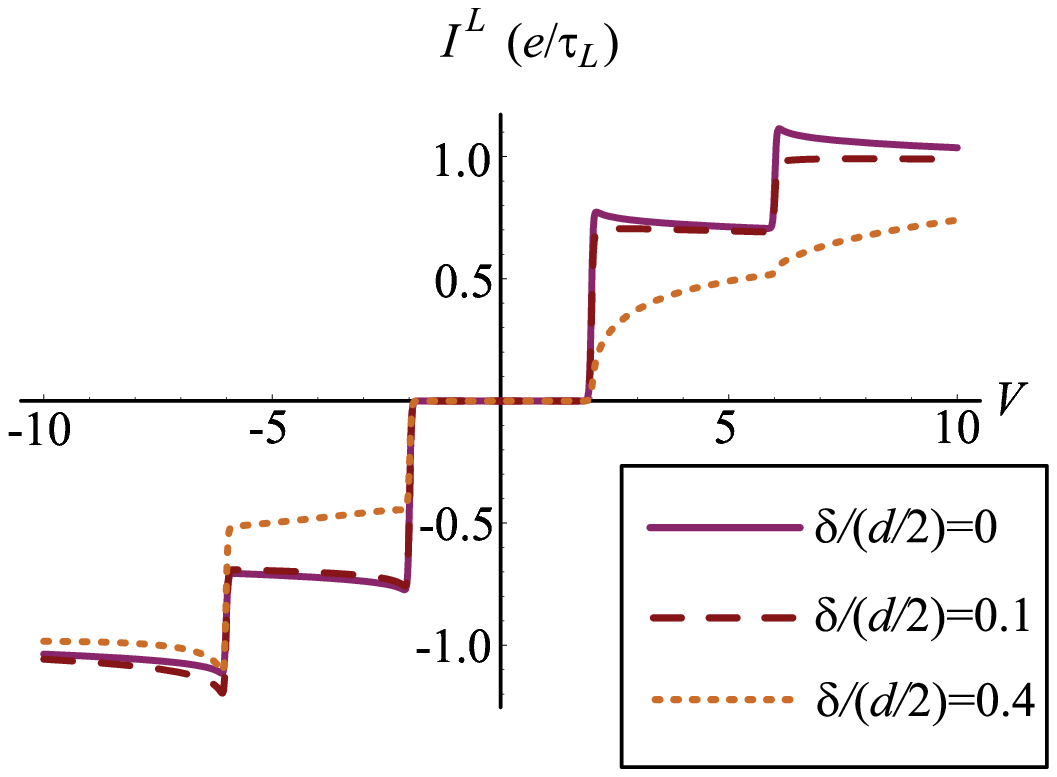}  
\end{array}$
\caption{(a) Current-voltage characteristics of a symmetric junction with nanotube leads for different inter-lead distances $d$.
(b) Current-voltage characteristics for different displacements $\delta\equiv d_L-d/2$ of the quantum dot. [Here $\delta=0$ represents a symmetric junction.]
We assume the same parameter values as used in Fig.~\ref{IVcurves}.}\label{position_asymmetry_dependence} 
\end{center}
\end{figure}

The two panels of Fig.~\ref{IVcurves} compare the current-voltage characteristics
of a quantum dot symmetrically side-coupled to two parallel Luttinger liquid leads 
with the inter-lead interactions included (upper panel) and neglected (lower panel).
We see that including inter-lead Coulomb interactions in our model enhances the electronic tunneling and changes the shape of the Coulomb-blockade steps qualitatively in comparison to the 
case of no inter-lead Coulomb interactions.
The enhanced tunneling, which manifests itself as regions of negative differential conductance,
is due to the Coulombic dot-lead interaction, Eq.~(\ref{Hcoul}),
as we discussed in Refs.~\onlinecite{Elste1} and \onlinecite{Elste2}.
However, in the regime of weak electron-electron interactions, 
this enhancement can be overcompensated by orthogonality effects, 
which results in an effective suppression of the current.
In general, the presence of inter-lead Coulomb interactions increases the total interaction strength 
and thus additionally enhances transport. 
The value of $Y_\alpha$ is thus smaller than it would be if inter-lead Coulomb interactions were ignored.

In the upper panel of Fig.~\ref{position_asymmetry_dependence} we demonstrate the effect of weakening 
the inter-lead interaction by moving the wires further apart.
Increasing $d$ weakens the strength of the
interaction and thus suppresses the tunneling of electrons. Increasing the inter-lead distance decouples the leads, causing the power-law 'divergences' explained above to vanish for sufficiently large inter-lead distances. Mathematically, the Luttinger exponents $Y_\alpha$ then assume values larger than unity.

Similarly, the lower panel of Fig.~\ref{position_asymmetry_dependence} demonstrates that breaking the spatial symmetry of the lead-dot-lead system can cause a suppression of 
the current near the Coulomb-blockade threshold, which would not occur 
in a quantum dot symmetrically side-coupled to the two leads.
The dependence of the Luttinger exponent $Y_L$ on the position of the dot relative to the leads
shows that a displacement of $\simeq 10\%$ already suffices to destroy the effect of enhanced electronic tunneling, since $Y_L$ then assumes values larger than unity.

\section{Conclusions}\label{conclusions}

We have studied transport through a quantum dot side-coupled to two parallel Luttinger liquid leads in the presence of a Coulombic dot-lead interaction.
The new physics considered in this paper is the inter-lead interaction.
The geometry is chosen to allow a bosonisation solution to the inter-lead interaction as
well as to the 
interaction between charge fluctuations on the dot and the dynamically generated image charge in the leads.
A master-equation approach that treats the hybridization perturbatively
has been applied to compute the tunneling current.

Our most important finding is that for symmetrical junctions the  inter-lead electron-electron interactions  enhance the electronic transport and for reasonable interaction strengths may even
change the overall sign of the Luttinger exponent $Y_\alpha-1$  such that a suppression ($Y_\alpha>1$) of the current near the Coulomb-blockade threshold  in the absence of inter-lead interactions turns into an enhancement ($Y_\alpha<1$) in the $I$-$V$ curves. The enhancement  manifests itself as regions of negative differential conductance. As the symmetry of the junction is broken by moving the dot closer to one lead, the effect decreases. 

An important topic for future research is extending our results to the case of an end-coupled dot,
where the inter-lead interaction breaks the mathematical translation invariance which is 
employed in the standard theory of edge effects in Luttinger liquids.

\acknowledgments

AJM acknowledges support from the National 
\mbox{Science} Foundation under grant DMR-1006282. FE acknowledges support 
from the Deut\-sche For\-schungs\-ge\-mein\-schaft.


\begin{thebibliography}{99}

\bibitem{Park}H. Park, J. Park, A. K. L. Lim, E. H. Anderson, A. P. Alivisatos, and P. L. McEuen, Nature \textbf{407}, 57 (2000).

\bibitem{Jo}M.-H. Jo, J. E. Grose, K. Baheti, M. M. Deshmukh, J. J. Sokol,
E. M. Rumberger, D. N. Hendrickson, J. R. Long, H. Park, and
D. C. Ralph, Nano Lett. \textbf{6}, 2014 (2006).

\bibitem{Galperin}M. Galperin, M. A. Ratner, and A. Nitzan, J. Phys.: Condens. Matter \textbf{19}, 103201 (2007).

\bibitem{Boulat08a}E. Boulat and H. Saleur, Phys. Rev. B {\bf 77}, 033409 (2008).

\bibitem{Boulat08b}E. Boulat, H. Saleur, and P. Schmitteckert, Phys.\ Rev.\ Lett.\ \textbf{101}, 140601 (2008). 

\bibitem{Goldstein09}M. Goldstein, Y. Weiss, and R. Berkovits, Europhys.\ Lett.\ \textbf{86}, 67012 (2009); \textit{ibid.}~Physica E \textbf{42}, 610 (2010); M. Goldstein, R. Berkovits, Phys.\ Rev.\ Lett.\ \textbf{104}, 106403 (2010).

\bibitem{Elste1}F. Elste, D. R. Reichman, and A. J. Millis, Phys.\ Rev.\ B \textbf{81}, 205413 (2010).

\bibitem{Elste2}F. Elste, D. R. Reichman, and A. J. Millis, arXiv:1010.2251

%\bibitem{Perfetto}E. Perfetto, G. Stefanucci, and M. Cini, Phys.\ Rev.\ Lett.\ \textbf{105}, 156802 (2010).

\bibitem{Fabrizio94}M. Fabrizio, A. O. Gogolin, and S. Scheidl, Phys.\ Rev.\ Lett.\ \textbf{72}, 2235 (1994).

\bibitem{Maurey}H. Maurey and T. Giamarchi, Phys.\ Rev.\ B \textbf{51}, 10833 (1995).

\bibitem{Lerner}Igor V. Lerner, Vladimir I. Yudson, and Igor V. Yurkevich, Phys.\ Rev.\ Lett.\ \textbf{100}, 256805 (2008).

\bibitem{Haldane81}F. D. M. Haldane, J.\ Phys.\ C: Solid State Phys.\ \textbf{14}, 2585 (1981).

\bibitem{Giamarchi}T. Giamarchi, \textit{Quantum Physics in One Dimension}, Oxford University Press, Oxford (2004). 

\bibitem{Luther}A. Luther and I. Peschel, Phys.\ Rev.\ B \textbf{9}, 2911 (1974).

\bibitem{Schotte}K. D. Schotte and U. Schotte, Phys.\ Rev.\ \textbf{182}, 479 (1969).

%\bibitem{Fabrizio97}M. Fabrizio and A. O. Gogolin, Phys.\ Rev.\ Lett.\ \textbf{78}, 4527 (1997).

%\bibitem{Furusaki97}A. Furusaki, Phys.\ Rev.\ B \textbf{56} 9352 (1997).

\end{thebibliography}
\end{document}